\documentclass[prb,aps,twocolumn,showpacs,preprintnumbers,amsmath,amssymb,10pt]{revtex4-1}
\usepackage[pdftex]{graphicx}% Include figure files
\usepackage{graphicx}
\usepackage{dcolumn}% Align table columns on decimal point
\usepackage{bm}% bold math

\usepackage{mathcomp}

\DeclareGraphicsExtensions{.pdf}   % Dateiendung fu"r PDF-Bilder
\usepackage[justification=raggedright,singlelinecheck=false]{caption}[2008/08/24]

\begin{document}

\title{DFT screened-exchange approach for investigating electronical properties of graphene-related materials}
%\titlerunning{Short title }

\author{Roland Gillen}\email{rg403@cam.ac.uk}
\author{John Robertson}
\affiliation{Department of Engineering, University of Cambridge, Cambridge CB3 0FA, United Kingdom}

%\authorrunning{Gillen et al.}

%\mail{e-mail
%  \textsf{rg403@cam.ac.uk}}

%\institute{%
%  \textsuperscript{1}\,Center of Advanced
%Photonics, Department of Engineering, University of Cambridge, Cambridge CB3 0FA, United Kingdom
%}

%\received{XXXX, revised XXXX, accepted XXXX} % do not change, will be filled in by the publisher
%\published{XXXX} % do not change, will be filled in by the publisher

%\pacs{ } % For example: 71.20.Ps

\date{\today}

\begin{abstract}
We present \textit{ab initio} calculations of the bandstructure of graphene and of short zigzag graphene nanoribbons by the screened-exchange-LDA method (sX-LDA) within the framework of density functional theory (DFT). The inclusion of non-local electron-electron interactions in this approach results in a renormalization of the electronic bandstructure and the Fermi velocity compared to calculations within local density approximation (LDA) gives 
good agreement with experiment. Similarly, the band gaps in zigzag nanoribbons (ZGNR) are widened by more than 200\%, being of similar magnitude than bandgaps from past studies based on quasiparticle bandstructures. 
We found a noticeable effect of non-local exchange on the spin-polarization of the electronic ground state of ZGNRs, compared to LDA and GGA-PW91 calculations.
\end{abstract}

\maketitle

\section{Introduction}
Since the first successful preparation of free-standing graphene, an isolated layer of graphite, in 2004\cite{novoselov04}, a considerable amount of both theoretical and experimental work have been employed on investigations of its  unique physical properties. Graphene is a zero-gap semiconductor where the $\pi$ valence band and the $\pi^*$ conduction band contact at the six Dirac points in the hexagonal graphene Brillouin zone. A peculiar property of graphene is the linear dispersion $E=\hbar v_F|\vec{k}|$ ($v_F$ is the Fermi velocity) of the $\pi$-bands near the K-points, resulting in effectively massless Dirac fermions\cite{novoselov-fermiongas,zhang-hall}. This property, predicted theoretically and confirmed experimentally, is appealing for theoretical scientists, as it allows in principle to study relativistic Dirac fermions by methods from condensed-matter physics. Similarly, the outstanding electronical properties turn graphene and its related materials, such as carbon nanotubes and graphene nanoribbons, to promising materials for application in nanoelectronics. For this reason, a thorough understanding of the electronic properties of graphene materials are significant.\\
A common and efficient method for the study of ground state properties in condensed-matter physics are \textit{ab initio} density functional theory calculations within the local density approximation (LDA) or generalized density approximation (GGA). Whereas those approaches, in all their simplicity, usually yield good results for structural properties they routinely underestimate electronical properties, \textit{e.g.} band gaps. Recent reports of experimental studies of charge carrier dynamics in graphene by use of angle resolved photoemission spectroscopy (ARPES)\cite{bostwick-ARPES,sprinkle-ARPES}, IR measurements\cite{orlita-fermivel}, magnetotransport measurements\cite{zhang-hall}, and scanning tunneling microscopy\cite{miller-fermivel}, revealed that common density functional theory calculations severely underestimate the Fermi velocity by 15
-20\%. This was attributed to prominent many-body effects in graphene such as non-local electron-electron and electron-phonon interaction, which are not accounted for in the LDA frame and lead to a renormalization of the Fermi velocity. Similarly, calculations within LDA underestimate the band gap sizes in both armchair and zigzag graphene nanoribbons\cite{son-energygaps} by 50-75\%, bad performances even for LDA calculations on semiconductors. Several authors have reported that self-energy corrections of the LDA bandstructures using many-body \textit{GW} approximations can amend the shortcomings of LDA. For the Fermi velocity, those corrections yield values that are in good agreement with experimental results\cite{trevisanutto-GW,attaccalite-GW}. However, while quasiparticle corrections produce accurate results, the considerable computational effort here is an undeniable disadvantage.\\
There are several attempts to improve on the shortcomings of LDA by including non-local expressions in the exchange-correlation functional. In case of the screened-exchange (sX-LDA) approximation, the electron-correlation functional is modeled by a blend of LDA and a screened Hartree-Fock exchange potential in order to account for electron-electron interactions to some extend but retaining the advantages of LDA\cite{byklein-SX,%Seidl-SX
xiong_sX}. It has been shown that the screened-exchange-method is successful in describing the band gaps of various materials with accuracy comparable to GW\cite{lee-gan, lee-hexa}.

Motivated by that success, we want to use this paper to report calculations on the bandstructure of graphene and zigzag graphene nanoribbons employing sX-LDA. We found that the calculated renormalization of the Fermi velocity in graphene due to electron-electron interactions compares well with the experimental values\cite{bostwick-ARPES,sprinkle-ARPES} and the ones from quasi-particle corrections\cite{trevisanutto-GW,attaccalite-GW}. Further, we show that the non-local exchange has a considerable effect on the spins-polarization and the electronic bandstructures of zigzag nanoribbons, which again compare well with quasiparticle bandstructures\cite{yang-GW}.

%Calculation
\section{Method}
Our study of the electronic bandstructures of graphene and zigzag graphene nanoribbons is based on pseudopotential density functional theory in the frame of the screened-exchange-LDA approximation, LDA and GGA in the PW91 form, and
was performed by use of the computational package \textit{CASTEP}\cite{castep}. Graphene was modeled by the common two-atomic Wigner-Seitz unit cell and periodic boundary conditions in the three spacial directions. We found that a distance of 6\,\AA\space between the periodic images is sufficient to minimize interlayer interactions and represent graphene as a 2D crystal. The action of the atomic core were described by standard normconserving pseudopotentials\cite{byklein-pseudo}, the valence electrons by plane waves with a cutoff energy of 750\,eV. Integrations in the reciprocal space were done by a Monkhorst-Pack grid\cite{monkhorst} of 9x9x1 equi-distant points in the Brillouin zone of graphene. We further optimized the atomic positions until the maximum interatomic force were lower than 0.001\,eV/\AA. %We want to note that the geometry optimization resulted in a lattice constant $a_0=2.52$\AA, which is by 2\% larger then the experimental lattice constant of 2.46\AA. Consequently, the interatomic distance in the unit cell is $d=1.69$\AA, \textit{i.e.} the screened-exchange approximation results in an underbinding of the crystal atoms. This, however, had no influence on the results of our calculations.\\
We modeled graphene nanoribbons by a rectangular unit cell being continued periodically along the nanoribbon axis and used an energy cutoff of 750\,eV and a k-point grid of 1x1x9 points along the Brillouin zone. The cell dimensions were chosen in a way to minimize interlayer interaction between the periodic images and to maintain the calculational effort at the same time. As for graphene, we found that a distance of 6\,\AA\space between periodic images is a sufficient compromise. The dangling bonds of the carbon atoms at the nanoribbons edges were passivated with hydrogen atoms in order to maintain the sp$^2$-hybridization. The geometries of all nanoribbons were fully optimized.

%Results and Discussion
\section{Results and Discussion}

%Fermi velocity
\subsection{Bandstructure of graphene}
\begin{table}[bt]
\caption{\label{tab:fermivelocities} Ab initio and experimental values of the Fermi velocity in the linear valence bands of graphene.}
\begin{tabular}{c|c}
Method&Fermi velocity $v_F$ (m/s)\\
\hline
LDA&$0.89\cdot 10^6$\\
GGA&$0.88\cdot 10^6$\\
sX-LDA&$1.16\cdot 10^6$\\
GW (CD integration)\cite{trevisanutto-GW}& $1.12\cdot 10^6$\\
GW (Random phase approximation)\cite{attaccalite-GW}& $1.25\cdot 10^6$\\
Experiment (ARPES)\cite{sprinkle-ARPES}& $1.0\pm 0.05 \cdot 10^6$\\
Experiment (IR)\cite{orlita-fermivel}& $1.02\pm 0.01 \cdot 10^6$\\
Experiment (magnetotransport)\cite{miller-fermivel}& $1.07\pm 0.01 \cdot 10^6$\\
Experiment (STM)\cite{zhang-hall}& $1.1 \cdot 10^6$\\
\end{tabular}
\end{table}

We used these parameters to calculate the electronic bandstructures of graphene in the frame of LDA, GGA and sX-LDA. 

\begin{figure*}[tbh]
\centering
\begin{minipage}{\columnwidth}
\includegraphics*[width=0.95\columnwidth]{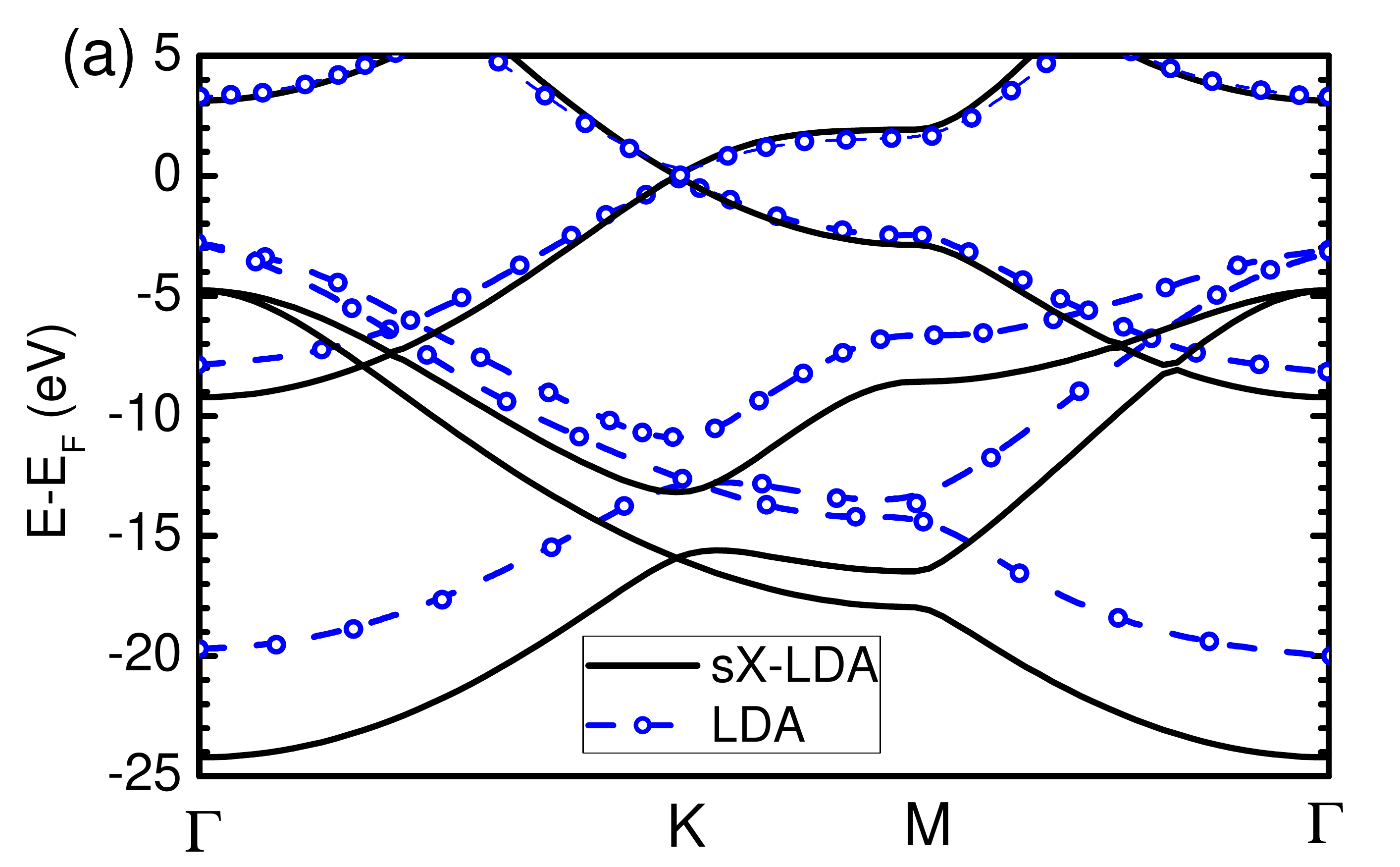}
\end{minipage}
\quad
\begin{minipage}{\columnwidth}
\includegraphics*[width=0.95\columnwidth]{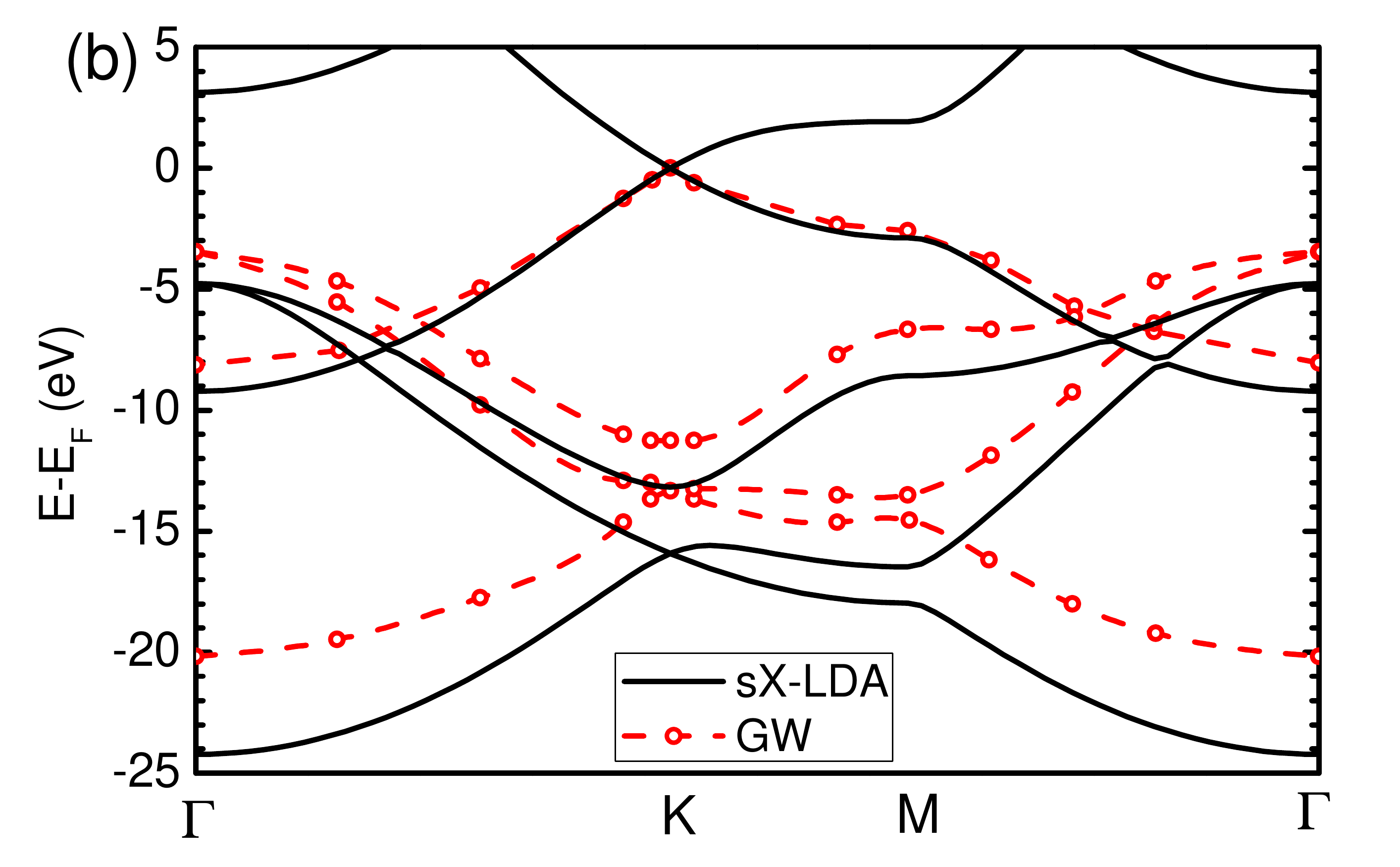}
\end{minipage}
\caption{\label{fig:bandstructures} (Color online) Comparison of the bandstructure of graphene within the screened-exchange approximation with (a) the bandstructure within the local density approximation (LDA) and (b) the quasi-particle bandstructure from self-energy corrections within the GW approximation, which was extracted from Ref.~\onlinecite{trevisanutto-GW}.}
\end{figure*}

%Figure 1 (a):
Figure~\ref{fig:bandstructures} (a) compares the valence bands of graphene from calculations within the local density approximation and with screened-exchange. There is a visible renormalization of the bandstructure due to the inclusion of non-local correlation effects. Screened-exchange results in a general shift of the three valence $\sigma$-bands to lower energies and that is most significant near the $\Gamma$-point. There, the two degenerate bands from sX-LDA have energies of -5\,eV compared to -3.3\,eV from LDA. The conduction bands are not affected as much from non-local exchange. This might correlate with studies by Lee \emph{et al.}\cite{lee-SX}, who argue that the improved band gaps from screened-exchange-LDA functionals in the materials they studied mainly originate from a marked downshift of the valence bands. This lowered energy of the valence bands compared to the Fermi energy results in a broadened band gap by $~$1.7\,eV at the $\Gamma$-point.
The $\pi$-bands from both calculated bandstructures coincide fairly well in the K-M part of the Brillouin zone.
In the $\Gamma$-K and $\Gamma$-M parts of the bandstructure, the non-local electron interaction results in a visibly increased slope for the sX-LDA $\pi$-band in respect to the $\pi$-band from LDA. This results in a renormalized Fermi velocity
\begin{equation}
v_F=\frac{1}{\hbar}\frac{\partial E}{\partial k}
\end{equation}
in the linear part of the $\pi$-bands from sX-LDA compared to LDA.
The Fermi velocity is a particularly interesting variable in semiconductors and conductors, as it is comparable to the velocity of the electrons that contribute to electric conduction. 
Table~\ref{tab:fermivelocities} shows a number of recently reported experimental values for the Fermi velocity, which are between $1.0\cdot 10^6$ and $1.1\cdot 10^6$\,m/s. For the bandstructure from LDA, we found a Fermi velocity of $8.9\cdot 10^5$\,m/s, which underestimates the reported experimental Fermi velocities by 11-19\% (depending on the experiment). Calculations within the generalized gradient approximation yield the same result, see Table~\ref{tab:fermivelocities}. In contrast, the steeper slope in the sX-LDA bandstructure results in a Fermi velocity of $1.16\cdot 10^{6}$\,m/s, being in good agreement with the values from magnetotransport\cite{miller-fermivel} and STM\cite{zhang-hall} measurements.

%Figure 1 (b)
Figure~\ref{fig:bandstructures} (b) shows the valence bands from calculations within sX-LDA, together with quasi-particle energies at selected k-points from GW calculations by Trevisanutto \emph{et al.}\cite{trevisanutto-GW}. The GW calculations build upon LDA wavefunctions and have been performed by use of a contour-deformation (CD) integration, which is known to provide the most accurate results in GW quasiparticle corrections. As with sX-LDA, the correlation effects induce a shift of the three $\sigma$-bands to lower energies, but the downshift is considerably weaker and of magnitude $<$1\,eV. Unfortunately, the authors did not report any data for the conduction band energies. Lee \emph{et al.}\cite{lee-SX}, however, found that the improved band gaps from GW result from a strong upshift of the conduction bands. The $\pi$ band from sX-LDA shows a very good agreement with the $\pi$-band in the quasiparticle bandstructure, except for deviations near the $\Gamma$-point. %In addition, the kink in the linear dispersion near the K-point, which is believed to stem from electron-plasmon effects\cite{bostwick-ARPES} and occurs also in GW bandstructures\cite{trevisanutto-GW,attaccalite-GW}, does not appear in our results. 
The value of the Fermi velocity was reported to be $1.12\cdot 10^6$\,m/s, \textit{i.e.}, only slightly closer to experimental values than our result. This encourages us to believe that the electronical properties of graphene and graphene-based materials can be described in good accuracy within the frame of the screened-exchange approximation.

%Nanoribbons
\subsection{Spin-induced band gaps in zigzag nanoribbons}
\begin{figure*}[tbh]
\centering
\begin{minipage}{\columnwidth}
\includegraphics*[width=0.95\columnwidth]{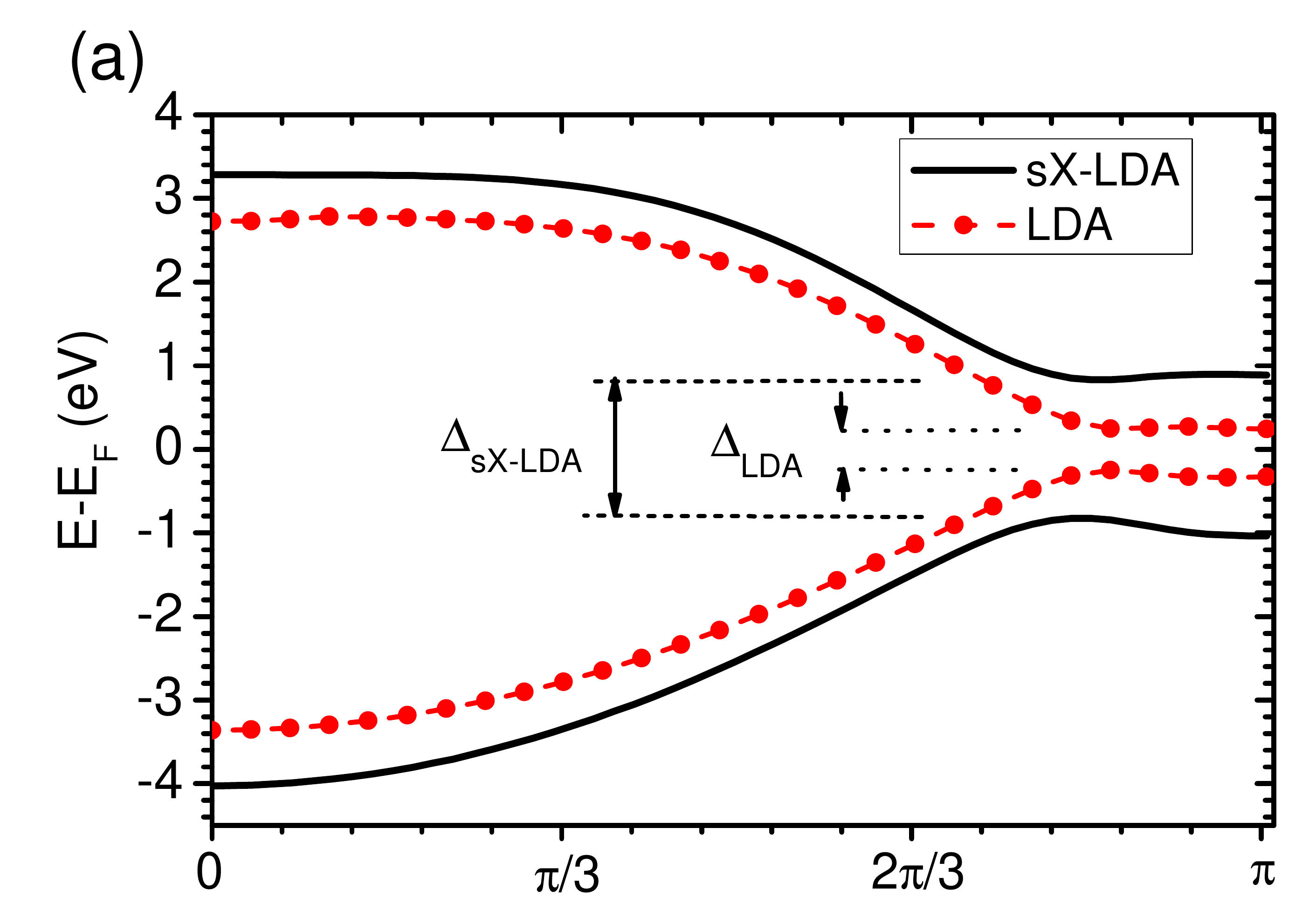}
%\caption{\label{fig:bandstruct-3ZGNR} Edge bands of a small zigzag nanoribbon (6 carbon atoms) from calculations employing screened-exchange (black lines) and local density approximation (red dots and broken lines).}
%\end{figure}
\end{minipage}
\quad
\begin{minipage}{\columnwidth}
%\begin{figure}[tbh]
%\centering
\includegraphics*[width=0.95\columnwidth]{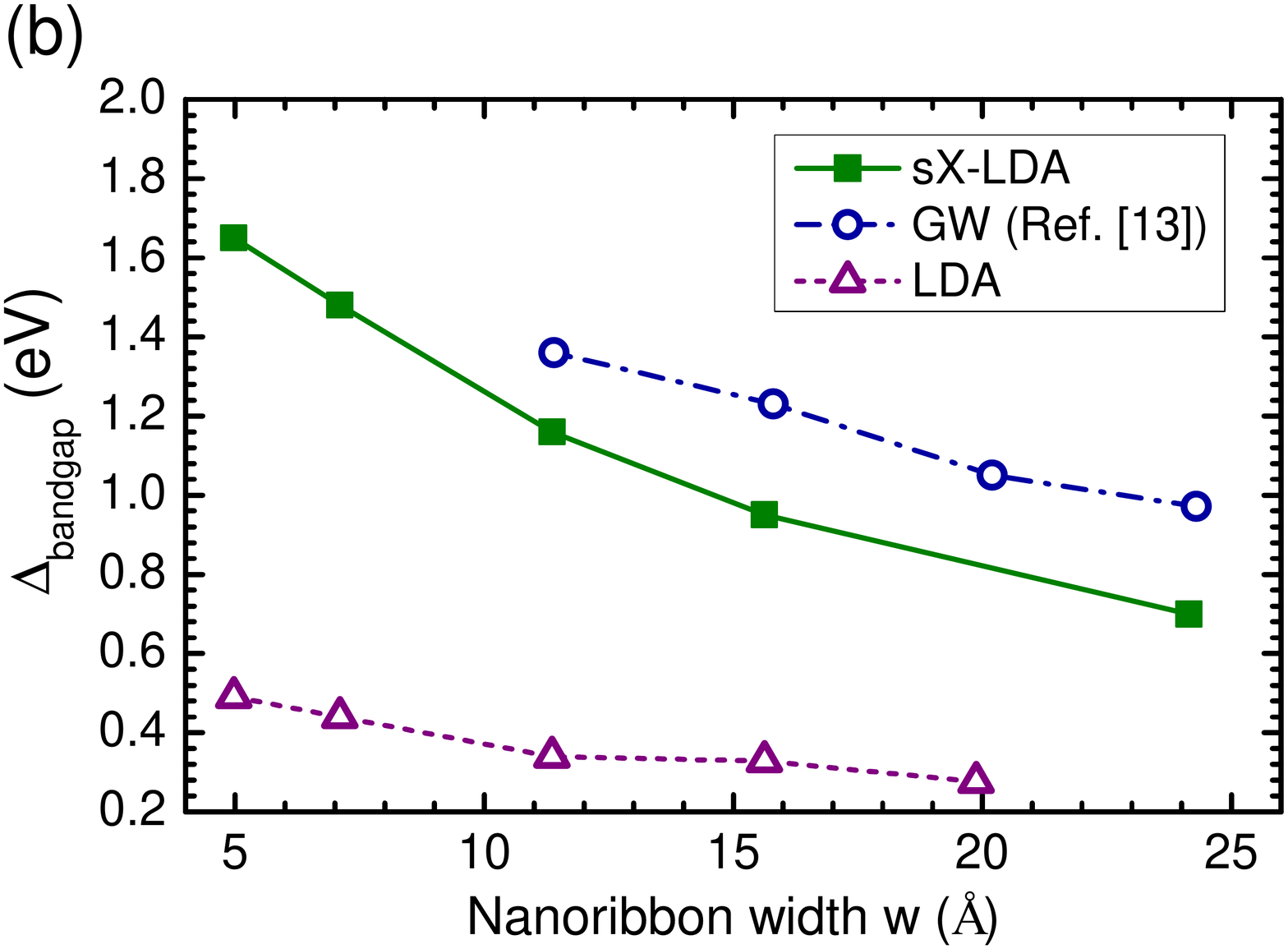}
%\caption{\label{fig:bandgaps-ZGNR} Comparison of the bandgap sizes of zigzag nanoribbons of various widths from sX-LDA (green circles), LDA+GW (red squares) and LDA (blue triangles) calculations. The GW results were taken from Ref.~\onlinecite{yang-GW}.}
\end{minipage}
\caption{\label{fig:bandgaps} (Color online) (a) Edge bands of a small zigzag nanoribbon (6 carbon atoms) from calculations employing screened-exchange (black lines) and local density approximation (red dots and broken lines). (b) Comparison of the bandgap sizes of zigzag nanoribbons of various widths from sX-LDA (green circles), LDA+GW (red squares) and LDA (blue triangles) calculations. The GW results were taken from Ref.~\onlinecite{yang-GW}.}
\end{figure*}

%In this section, we discuss our results for the bandgap sizes of zigzag graphene nanoribbons. 
The first investigations of the electronical strutures of graphene nanoribbons were based on simple zonefolding arguments, neglecting effects from the nanoribbon edges. In this approximation, one third of all nanoribbons with armchair edges and all zigzag nanoribbons should be metallic or zero-gap semiconductors\cite{fujita96,ezawa-GNR}, similar to the situation in carbon nanotubes. More elaborate investigations employing density functional theory predicted that the existence of a large ratio of edge to 'bulk' atoms in carbon nanoribbons indeed has a prominent effect on the electronical structure. For zigzag nanoribbons, the deciding factor are edge-localized electron states, which possess a large density of states at the Fermi energy. Using a Hubbard model, Fujita \emph{et al.}\cite{fujita96} showed that the resulting Fermi instability leads to the formation of an antiferromagnetic ground state in monohydrogenated nanoribbons in contrast to the dielectric graphene. This spin-polarized ground state involves the opening of a direct bandgap as the system wants to lower the density of states near the Fermi level and remove the instability\cite{pisani-GNR}.

Investigations employing LDA predict that the bandgap size is lower than 0.5\,eV for all zigzag nanoribbons and scales antiproportionally with the nanoribbon width due to the decrease of spin-interaction between the atoms at opposite edges\cite{son-energygaps}. It might be expected from this interaction that nonlocal components of the electron exchange are of importance for the bandgap size and that LDA, neglecting nonlocal exchange, is not sufficient to fully describe the electronical properties of zigzag nanoribbons. Indeed, our calculations exhibit marked changes in the bandgap size due to the inclusion of nonlocal exchange. Figure~\ref{fig:bandgaps}~(a) shows the electron bands of the edge states of a small zigzag nanoribbon from LDA and sX-LDA calculations. While both bandstructures qualitatively exhibit similar dispersion, the bandgap near $k$=$\frac{2\pi}{a}$ is widened by more than 200\% from a value of ~0.5\,eV in LDA to 1.65\,eV in sX-LDA. Our calculated band gap energies are comparable to those obtained from GW quasiparticle corrections on LDA wavefunctions by Yang et al.\cite{yang-GW}, see Fig.~\ref{fig:bandgaps}~(b), but the quasiparticle bandgaps are consistently ~0.2\,eV larger than our values. Another study, employing the B3LYP hybrid potential, reported for an 8-ZGNR (16 carbon atoms, w$\approx$ 16\AA) a bandgap size of 1.34\,eV\cite{rudberg-GNR}, compared to 1.02\,eV from sX-LDA and 1.23\,eV from GW. So far, no experimental data exists for comparison of those calculated band gaps sizes with real values, however, the similar results from different non-local approaches suggests that the small values from LDA are indeed a severe underestimation of the real bandgap energies.
There are two different factors for the size of the band gap. One is the quantum confinement, which induces a inversely proportional width dependence band gap energies, the other is the separation of the valence and the conduction band by formation of a spin-polarized ground state. 
To estimate the influence of the spins on the band gap size, we fitted a function
\begin{equation}
E_g=\frac{A}{w + \Delta}
\end{equation}
to the sX-LDA band gap energies. $\Delta$ is the deviation length from the ideal quantum confinement law $E_g\propto \frac{1}{w}$ and is an indicator for the effect of the spin interaction. $\Delta=9.3$\,\AA\space gives good agreement with our calculated values. Yang \emph{et al.} reported a value of $\Delta=16$\,\AA\space, which might imply a stronger contribution of the interaction of the electrons with the surrounding system in their GW calculations.\\
\begin{figure}[bh]
\centering
\includegraphics*[viewport=0 540 570 835,width=0.95\columnwidth]{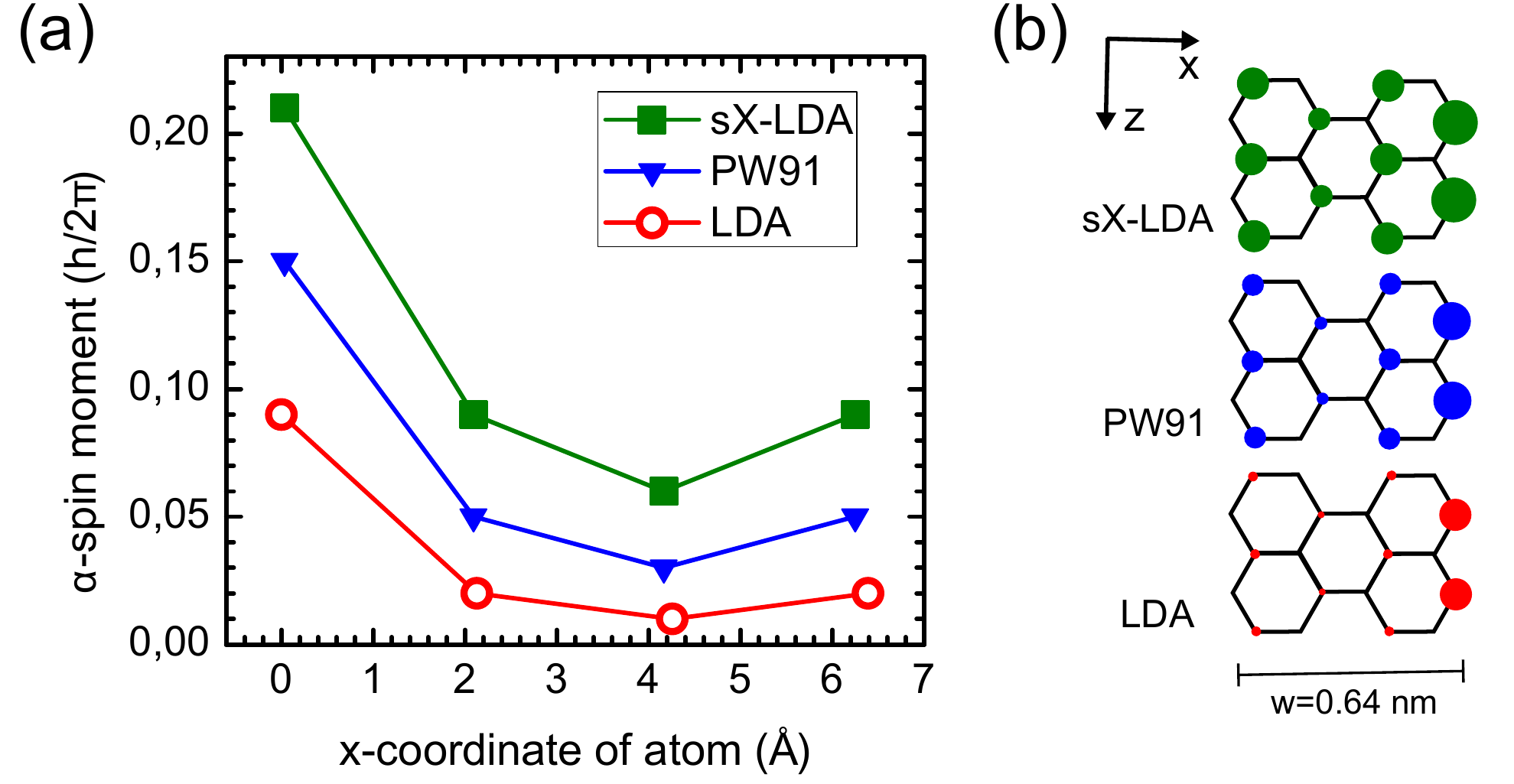}
\caption{\label{fig:spin-gs} (Color online) (a) Comparison of $\alpha$ spin momenta of a 4ZGNR (eight atoms per unit cell) from calculations employing LDA (red circles). (b) Graphical representation of the spin momenta from (a) on the nanoribbon atoms. The ratio of the areas of the circles representing the spin momenta is the same as the ratio between the absolute values of the momenta.}
\end{figure}

We obtained the spin polarization of the electronic ground states of our investigated nanoribbons by optimization of the atomic positions and the electronic spins. Starting from a non-spin-polarized geometry, these optimizations resulted in anti-ferromagnetic ground states all our calculations within LDA, GGA-PW91 and sX-LDA.
Figure~\ref{fig:spin-gs} (a) shows the momenta of the $\alpha$-spin direction in case of a 4-ZGNR, a nanoribbon with 8 carbon atoms per unit cell. The ground state from LDA calculations exhibits a comparatively weak spin-polarization, which is strongly localized at the nanoribbon edges, see Fig.~\ref{fig:spin-gs} (b). The calculated spin momentum at the edges is 0.09\,$\hbar$ and rapidly decreases to a almost negligible value of 0.01\,$\hbar$ at the nanoribbon 'bulk' atoms. This corresponds to a decrease of 88\% over the distance of second-nearest neighbours. While the spin-momenta from sX-LDA still show a noticeable localization at the edges, the spin-polarization over the whole unit cell of the 4-ZGNR is more balanced. The spin momentum at the edges is 0.21\,$\hbar$, more than 2.5x the value from LDA, and decreases by 71\% into the nanoribbon 'bulk'. As expected, GGA improves on the magnetic properties of the material. We found that that the spin momenta from PW91 are quite in-between the values from LDA and sX-LDA, falling from 0.15\,$\hbar$ at the edges down to a value of 0.03\,$\hbar$, \textit{i.e.} a drop of about 80\%. The same results hold for the momenta of the $\beta$ spins, but from the opposite edge and with negative values, and for wider nanoribbons. It is evident from the degree of spin-polarization and the changed band gap energies that the nonlocal component of the electron exchange indeed has a significant effect on the electronic properties of zigzag nanoribbons.

\section{Conclusion}
In conclusion, we used the screened-exchange (sX-LDA) approximation to calculate the electronic bandstructure of graphene and zigzag graphene nanoribbons and compared the results with the bandstructures from LDA and the quasiparticle bandstructures from the GW approach. The introduced electron-electron interaction in sX-LDA leads to a renormalization of the bandstructure and the Fermi velocity in the linear part of the $\pi$-band in graphene, as was found in experiments and from GW calculations. We report that the Fermi velocity from sX-LDA is comparable to the value from GW calculations and in good agreement with experimental values. For zigzag nanoribbons, our calculations confirmed the significance of the nonlocal part of the electron exchange interaction for the degree of spin-polarization and for the size of the spin-induced bandgap in the electronic bandstructure. The nonlocal exchange results in a larger bandgap compared with LDA and comparable to that from GW. The spin momenta in the electronic ground state from sX-LDA calculations are noticeably higher and more balanced over the whole unit cell of graphene nanoribbons compared to results from LDA and GGA-PW91. As a result, we are confident that the screened-exchange-LDA approach is a useful alternative method for the study of electronic properties in graphene-related materials.

\end{document}